\documentclass[11p]{article}
\pdfoutput=1
\usepackage{hyperref} 
\DeclareUnicodeCharacter{223C}{\textasciitilde}
\usepackage{braket}
\usepackage[utf8]{inputenc}
\usepackage[T1]{fontenc}
\usepackage{ amssymb }
\usepackage[center]{caption}
\usepackage{graphicx,epsfig} 
\usepackage[english]{babel}
\usepackage{lscape} 
\usepackage{mathtools} 
\usepackage{amsmath}
\usepackage{amsthm}
\usepackage{amsfonts} 
\usepackage{fancyhdr} 
\usepackage{chngcntr}
\usepackage{xcolor}
\usepackage{soul}
\usepackage{float} 
\usepackage{subcaption}
\usepackage[labelsep=period]{caption}
\usepackage{multirow}
\usepackage{physics}
\usepackage{dsfont}
\usepackage{MnSymbol}
\usepackage{bbm}
\usepackage{tikz}
\usetikzlibrary{arrows.meta}
\usepackage{tikz-cd}
\usetikzlibrary{positioning}
\usepackage[a4paper]{geometry}
\date{}
\author{Sami Calvo}
\usepackage{fancyhdr,lipsum}
\DeclareCaptionFormat{custom}{\textbf{#1#2}\textit{#3}}

\usepackage[textfont=normalfont]{caption}

\usepackage{titlesec}
\titleformat{\section}{\Large\bfseries\filcenter}{\thesection.}{1em}{}
\titleformat{\subsection}{\Large\bfseries\filcenter}{\thesubsection.}{1em}{}
\titleformat{\subsubsection}{\Large\bfseries\filcenter}{\thesubsubsection.}{1em}{}

\date{}

\newtheorem{proposition}{Proposition}
\newtheorem{theorem}{Theorem}
\newtheorem{lemma}{Lemma}

\newtheorem*{remark}{Remark}

\begin{document}
\theoremstyle{definition}
\newtheorem{definition}{Definition}[section]
\title{On the Stochastic-Quantum Correspondence}
\author{Sami Calvo\footnote{Physics Faculty, University of Barcelona, Barcelona, 08028, Spain. Email: samicalvo@gmail.com}}
\date{}

\maketitle
\begin{abstract}
    \noindent This paper aims to first explain, somewhat more clearly, the Stochastic-Quantum correspondence put forward in by Barandes in 2023. Specifically, the quantum-mechanical bra-ket notation is used, illuminating some results of previous results. With this, we prove the six axioms of textbook quantum mechanics from a single axiom: every physical system evolves according to a, generally indivisible, stochastic law. Afterwards, we generalise the treatment to continuous bases, which showcases a problem with them, indicating that space (and other physical variables) may be discrete in nature. Some concrete examples are also given, including the generalisation to classical and quantum fields. Then, we treat some practical issues of this new stochastic approach, regarding the solving of problems in physics, which turns out to still be most tractable in the traditional way. Finally,  we explain the classical limit, where a system of many particles is found to behave classically according to Newton's second law. Along with that, we present a way of solving the measurement problem, characterising what is an environment and a measuring device and explaining how the wavefunction collapse comes about. Specifically, it is found that what distinguishes an environment is its number of degrees of freedom, while a measuring device is a low-entropy type of environment.
\end{abstract}
\section{ Introduction}
Since the birth of quantum mechanics, in the early 20th century, many attempts have been made to find philosophically satisfactory interpretations of the theory \cite{Bell,Bohm,Everett, Zurek,Ghirardi,Spekkens, Schlosshauer, Wallace, Spekkens2}, including some (divisible) stochastic proposals \cite{Fenyes, Nelson4, Bohm2, Ballentine2, Peña}. However, all of these interpretations suffer from diverse philosophical and scientific problems: either they can't completely reproduce experimentally verified results (called empirical adequacy) and/or they offer completely foreign world-views to everyday experience. However, a new approach has recently emerged that is both capable of reproducing experimental results and offers a satisfactory and unified world picture compatible with classical (everyday) physics. This new approach is called the Indivisible Stochastic approach \cite{Correspondence,Theorem}, embodied in the Stochastic-Quantum Correspondence\footnote{See footnote 6.}. We give here a brief summary of this correspondence and its components:
\\
\begin{itemize}

    \item \textbf{Stochastic process}\\\\
    One starts with a configuration space $\mathcal{C}$ of cardinality $C$, which, in the beginning, we will assume to be finite ($C\in\mathbb{N}$). This configuration space models some physical system, and each configuration corresponds to a physical state of the system. Then, a  stochastic law $\Gamma$ is postulated, which tells us the probability that the system is in state $i$ at time $t$ given that the system was in state $j$ at time $t=0$:
\begin{equation}
    \Gamma_{ij}(t)=p(i,t|j,0).
\end{equation}
Since the values $\Gamma$ can take are always positive, because they are probabilities, we can equate each entry of $\Gamma$:\footnote{Note that this is not a matrix multiplication, but an equality for individual entries. In the language of \cite{Correspondence,Theorem}, it is a Schur-Hadamard product.} $\Gamma_{ij}=|\Theta_{ij}|^2\ge0$, for some complex-valued $\Theta$. As it is proven in \cite{Theorem}, it turns out that $\Gamma$ can always be expressed in a unistochastic form, that is, $\Theta$ is a unitary matrix (and from now on, it is called $U$ instead of $\Theta$). \\\\
Since probabilities add to one, we have the condition
\begin{equation}
    \sum_i\Gamma_{ij}=1,
\end{equation}
which, together with the non-negativity of its entries, makes $\Gamma$ a \textit{stochastic matrix}. Then, given the initial probabilities for the system being at some state $i$ we have the marginalisation formula:
\begin{equation}
    p(i,t)=\sum_j\Gamma_{ij}(t)p(j,t=0)\rightarrow  p(t)=\Gamma(t)p(0).
\end{equation}
A necessary condition for a process to be \textit{divisible} at a time $t'>t$ is that the following factorisation of $\Gamma$ is possible
\begin{equation}
\label{factorisation}
    \Gamma(t)=\Gamma(t'\to t)\Gamma(t'),
\end{equation}
for some \textit{stochastic matrix} $\Gamma(t'\to t)$. That means that we can \textit{divide} our time evolution into an intermediate time step $t'$. This is equivalent to $\Gamma$ being such that the following diagram commutes:
\begin{figure}[H]
\centering
\begin{tikzcd}[
    cells={nodes={font=\large}},
    every label/.append style={font=\large},
    column sep=5em,
    row sep=4em
]
\text{Time } t=0
  \arrow[r, "\Gamma(t')"]
  \arrow[from=1-1, to=2-2, "\Gamma(t)"']
&
\text{Time } t=t'
  \arrow[d, "\Gamma(t' \to t)"]
\\
&
\text{Time } t=t
\end{tikzcd}

\caption{The commutation of this diagram is a necessary condition for the existence of a division event at $t=t'$.}
\label{fig:gamma-diagram}
\end{figure}

Then, the \textit{sufficient} condition is that the theory indeed gives this transition matrix. Specifically, it must be the case that
\begin{equation}
    \Gamma_{ij}(t'\to t)=|U_{ij}(t'\to t)|^2,
\end{equation}
with $U_{ij}(t'\to t)$ necessarily given by
\begin{equation}
\label{unitary transition}
    U_{ij}(t'\to t)=\sum_kU_{ik}(t)U^{-1}_{kj}(t').
\end{equation}
That is because if $U(t'\to t)$ is to be understood as a \textit{transition} amplitude, it must satisfy the following equality
\begin{equation}
    U(t'\to t)U(t')=U(t),
\end{equation}
which is exactly equivalent to Eq.~\eqref{unitary transition}. A more \textit {categorical} (in the sense of \textit{Category Theory}) and compact way of expressing this is to say that $\Gamma(t'\to t)$ must satisfy the following diagram:

\begin{figure}[H]
\centering
\begin{tikzpicture}
\node (top) {
\begin{tikzcd}[
    cells={nodes={font=\large}},
    every label/.append style={font=\large},
    column sep=5em,
    row sep=4em
]
\text{Time } t=0
  \arrow[r, "U(t')"]
  \arrow[dr, "U(t)"'] & \text{Time } t=t' \arrow[d, "U(t' \to t)"] \\
  & \text{Time } t=t
\end{tikzcd}
};

\node (bottom) [below=4em of top] {
\begin{tikzcd}[
    cells={nodes={font=\large}},
    every label/.append style={font=\large},
    column sep=5em,
    row sep=4em
]
\text{Time } t=0
  \arrow[r, "\Gamma(t')"]
  \arrow[dr, "\Gamma(t)"'] & \text{Time } t=t' \arrow[d, "\Gamma(t' \to t)"] \\
  & \text{Time } t=t
\end{tikzcd}
};

\draw[-{Latex}] ([xshift=-3em,yshift=3em]top.south west) to[out=250, in=110] node[left, font=\large] {$|\cdot|^2$} ([xshift=-3em,yshift=-1em]bottom.north west);

\draw[-{Latex}] ([xshift=3em,yshift=3em]top.south east) to[out=290, in=70] node[right, font=\large] {$|\cdot|^2$} ([xshift=3em,yshift=-1em]bottom.north east);

\end{tikzpicture}
    \caption{Necessary and sufficient condition for $t'$ to be a division event. Specifically, we want $\Gamma(t)=\Gamma(t'\to t)\Gamma(t')$ with $\Gamma_{ij}(t'\to t)=|U_{ij}(t'\to t)|^2$.}
    \label{fig:placeholder}
\end{figure}
Note that if $\Gamma$ is invertible, this condition is automatically satisfied (and that is why $U$ being unitary automatically guarantees it). Conversely, if the factorisation \eqref{factorisation} is not possible ---meaning $\Gamma$ is not invertible---, we say that the evolution is \textit{indivisible} at $t'$. The Stochastic-Quantum Correspondence, in order to reproduce Quantum Mechanics, is grounded on the evolution being indivisible in general.

\item \textbf{Hilbert space}\\\\
The configuration space, as already mentioned, models physical states. Therefore, every possible state is an eigenstate of a self-adjoint operator in the Hilbert space (with possibly more than one option for this operator)\footnote{Given any \textit{closed} subset $S$ of $\mathbb{R}$, we can always find a self-adjoint operator whose spectrum is precisely $S$. This can be seen by explicitly constructing a multiplication operator with eigenvalues which form a dense subset of $S$, and then the spectrum is the closure of these. However, many \textit{different} operators can have the same spectrum.}. This set of eigenstates forms a privileged orthonormal basis, called the \textit{configuration basis}, which in turn uniquely (up to isomorphism) determines the Hilbert space (all Hilbert spaces with the same cardinality are isomorphic to one another). We call this basis by $\{ \ket{i}_{i\in\mathcal{C}}$. With this, we make the following definitions:
\begin{enumerate}
    \item A projector: $P^{(i)}\coloneq\ket{i}\bra{i}$.

    \item A density operator: $\rho(t)\coloneq U(t)\left(\sum_i p_i(0)P_i\right)U^\dagger(t)$, which satisfies $\textnormal{tr}(\rho)=1$ and is self-adjoint. Then, the marginalisation formula becomes $p_i(t)=\textnormal{tr}( P_i\rho(t))$.
    
    \item A state vector\footnote{In this paper, the word ``wavefunction'' and ``state vector'' are used interchangeably, although the former is a particular basis representation, the position basis, of the latter.}: $\Psi^{(j)}_{i}\coloneq\sum_k U_{ik}\ket{j}_k$, where the $j$ index indicates a labelling whereas the $k$ one indicates components. 
    
    \item A Hamiltonian (as the generator of the unitary time evolution operator)\footnote{Setting $\hbar=1$.}: $\hat{\mathcal{H}}\coloneq i\frac{\partial U(t)}{\partial t}U^\dagger(t)$. 
\end{enumerate}
\end{itemize}
As we will shortly see, the connection between the two sides ---the Stochastic Process and the Hilbert Space--- is possible due to the representation of the $\Gamma$ matrix in terms of the unitary operator $U$, $\Gamma_{ij}=|U_{ij}|^2$, which will become the time-evolution operator of the wavefunction.
\vspace{0.5cm}

\section{ Axiomatic Theory}
Quantum mechanics can be axiomatically developed, as it was historically first done in \cite{Dirac, Neumann} and recently in, for example, \cite{Hardy, Ariano}. The six axioms that are normally used are:
\begin{enumerate}
    \item Every state is characterised by a vector in a Hilbert space, $\ket{\psi}$.

    \item Every observable corresponds to a self-adjoint operator on the Hilbert space.

    \item \textbf{(The Measurement Axiom)} After a measurement is performed, the state vector collapses to one of the eigenvectors of the corresponding operator. 
    
    \item \textbf{(Born's rule)} The probability of the collapse to the eigenvector $\ket{i}$ is given by $p_i=|\braket{i}{\psi}|^2$.

    \item \textbf{(Superposition Principle)} If a system can be in two states $\ket{\psi_1}$ and $\ket{\psi_2}$, it can also be in any arbitrary linear combination $c_1\ket{\psi_1}+c_2\ket{\psi_2}, c_i\in\mathbb{C}$.

    \item \textbf{(Schrödinger's Equation)} The time evolution of the state vector is given by $i\partial_t\ket{\psi}=\hat{\mathcal{H}}\ket{\psi}$, where $\hat{\mathcal{H}}$ is the Hamiltonian.
\end{enumerate}
From these axioms, the whole of quantum mechanics is derived. However, they have some major (philosophical) problems. The most obvious one is that they seem very arbitrary and far away from any physical intuition. Why Hilbert spaces? Where do Born's rule and Schrödinger's equation come from? What constitutes a measurement and how does it collapse the wave function? What is the physical interpretation of all this?
\\\\
In the stochastic approach, on the other hand, we only have the following axiom:
\begin{itemize}
    \item \textbf{Stochastic Axiom}: \textit{The evolution of every physical system is governed by a generalised stochastic process}
\end{itemize}
The term ``generalised stochastic process'' corresponds to the ``stochastic process'' introduced in the first section that, \textit{in general}, will be \textit{indivisible}. Note that a \textit{deterministic} system is a type of generalised stochastic system. In particular, it is divisible and $\Gamma$ is a permutation matrix (every column is full of zeros except for a one in one of its entries). Also, a generalised stochastic system can model a plethora of different systems: a particle in different positions in space, different momenta, different energies, field configurations, qubits in a quantum computer, etc., as we will later see. This is not the first stochastic approach to quantum theory \cite{Nelson, Nelson2, Nelson3, Nelson4, Guerra, Davidson, Cufaro, Yasue}; but it is the first (following \cite{Correspondence}) \textit{indivisible} approach with a satisfactory underlying philosophy (ontology) which is based on quantum theory, as opposed to classical dynamics.
\begin{theorem}
    \text{(The Stochastic-Quantum Correspondence)}\footnote{Here the terminology differs from \cite{Theorem}, where the ``Stochastic-Quantum Correspondence'' is the name given to the Theorem that any stochastic matrix is unistochastic, diluting (exapnding) the underlying Hilbert space if necessary.}. The six traditional, textbook quantum mechanics axioms can be derived from the Stochastic Axiom. 
\end{theorem}
\begin{proof}

    \begin{itemize}
    
        \item \textbf{Born's rule}. It immediately comes from $\Gamma$ being a \textit{transition} matrix, if we define $\ket{\psi}\coloneq U(t)\ket{j}$ (\textit{i.e.}, $\ket{\psi}$ is the time evolution of a basis element $\ket{j}$):
        \begin{equation}
            \Gamma_{ij}=|\bra{i}U(t)\ket{j}|^2=|\braket{i}{\psi}|^2\equiv |\psi_i|^2.
        \end{equation}

        \item \textbf{Axiom 1}. From the previous rule, every physical system is characterised by a state vector $\ket{\psi}$. That is, given the Born rule -- which we have from the system being governed by the \textit{unistochastic}\footnote{As proven in \cite{Theorem}, every stochastic matrix is unistochastic, at least in a dilated Hilbert space.} matrix $\Gamma$ -- we trivially find that the system can be fully described by the state vector $\ket{\psi}$

        \item \textbf{Schrödinger's equation}. From the definition of the state vector, it is immediate that it follows a Hamilton-type equation: 
        \begin{equation}
            i\partial_t\ket{\psi_j}=i(\partial_tU(t))\ket{j}=i(\partial_tU(t))UU^\dagger\ket{j}=(i\partial_tU(t)U^\dagger(t))\ket{\psi}=\hat{\mathcal{H}}\ket{\psi},
        \end{equation}
        where $\hat{\mathcal{H}}\coloneq i(\partial_tU(t))U^\dagger(t)$\footnote{The reader may think that the fact that $\ket{\psi}$ follows Schrödinger's equation is just a matter of a mere appropriate definition of $\hat{\mathcal{H}}$. However, $\ket{\psi}$ evolves with a unitary operator if and only if $\hat{\mathcal{H}}= i(\partial_tU(t))U^\dagger(t)$, and the \textit{unitary} time evolution comes from $\Gamma$ being unistochastic.}. The imaginary unit comes from $U$ being unitary and $\hat{\mathcal{H}}$ being self-adjoint. Indeed, by Stone's theorem and its inverse, $U(t)$ is unitary if and only if $\hat{\mathcal{H}}$ is self-adjoint. As we will see in the next section, the unitarity of $U(t)$ is not a consequence of probability conservation, as usually stated, but a necessary feature stemming from $\Gamma$ being \textit{unistochastic}. Moreover, the self-adjointness of $\hat{\mathcal{H}}$ can be seen as an experimental consequence (we only measure \textit{real} energies) or as an analogy to classical mechanics, where the Hamiltonian (normally) represents the energy of the system. 
        
        \item \textbf{Superposition principle}. It is derived in \cite{Correspondence} (section IV. D)--- although not quite explicitly stated there --- and the interference formula derived there is, together with the Born rule, equivalent to the superposition principle: if $\ket{\psi}$ is a sum of two other states, you get an interference when mod-squaring its (weighted) sum. Vice versa, if you get an interference term, it is because the mod-squaring was applied to a (weighted) sum:
        \begin{equation}
            |\psi_1+\psi_2|²= |\psi_1|²+|\psi_2|²+\text{Re}(\psi_1^*\psi_2)+\text{Re}(\psi_2^*\psi_1) 
        \end{equation}

        \item \textbf{Measurement axiom}. Its proof is relegated to Section 6.1. What we can already say, however, is that the Stochastic-Quantum correspondence has a very different philosophical stance. Instead of having a system that \textit{is} in a mixture of ``definite'' (eigenvector) states until the system ``collapses'' to one such state, we have a system that always is in a definite configuration, and the corresponding wavefunction is the one that collapses. In particular, the state is always in one of its possible configurations, which may have other associated properties (like position having momentum or energy associated to it), whereas the corresponding wavefunction may be in a state that is not an eigenvector of (almost) any operator.

        \item \textbf{Axiom 2}. The fact that every observable corresponds to a self-adjoint operator in a Hilbert space is a much more prosaic issue than in textbook quantum theory. Since the system now always has a definite configuration, we can always measure its properties without any problems. Then, as it is shown in section V.B of \cite{Correspondence}, on the Hilbert space side that corresponds to a self-adjoint operator acting on the system's state vector. Specifically, the self-adjointness property is crucial for being able to decompose the operator into its projectors via the spectral theorem. Moreover, self-adjointness guarantees that its eigenvalues are real.
    \end{itemize}
\end{proof}
\begin{remark}
    It has been proven that the system's wavefunction follows a Hamiltonian-type equation; however, we don't get the exact \textit{form} of the Hamiltonian only from the stochastic approach. This, on the one hand, allows for many different Hamiltonians, which is a good thing. But, on the other hand, it necessitates an outside imposition of the analytical expression of $\hat{\mathcal{H}}$. For example, one could, on symmetry grounds, impose that $\ket{\psi}$ satisfies a certain equation, like Dirac's or Klein-Gordon's, but that is, \textit{a priori}, not given by the stochastic approach.
\end{remark}
\noindent With this, we have a correspondence between a stochastic process and a quantum system. The configuration space uniquely (up to isomorphism) determines the Hilbert space and non-uniquely\footnote{Any basis consisting of the eigenvectors of a self-adjoint operator whose spectrum is $\mathcal{C}$ is valid.} determines the configuration basis, whereas the stochastic law $\Gamma$ allows for different Hamiltonians. Conversely, a given Hilbert space, together with an orthonormal basis,  uniquely determines the configuration space, and the Hamiltonian uniquely determines the stochastic law. Therefore, we have a one-to-many\footnote{If no specific basis is specified, the correspondence is many-to-many} correspondence between a stochastic process and a triple consisting of a Hilbert space, an orthonormal basis and a Hamiltonian:
\begin{equation}
    \textnormal{(}\Gamma,\mathcal{C}\textnormal{)}\overset{one-to-many}{\longleftrightarrow}(\mathcal{H}, \hat{\mathcal{H}},\{ \ket{\varphi^{(i)}}\}_{i\in\mathcal{C}}).
\end{equation}

\vspace{0.5cm}

\section{ Generalisation to the continuous case}
In \cite{Correspondence, Theorem}, all the treatment is done with a discrete, finite configuration space. In this case, we have that $\Gamma$ and $U$ are matrices that depend on time. The generalisation to the discrete, \textit{infinite} case is trivial: the sums in the above formulas now stretch up to infinity, and the rest is basically the same.
\\\\
However, in basic quantum mechanics, one normally studies a particle moving in space under some potential. Therefore, the configuration basis is the \textit{position basis}-- which is continuous-- and the corresponding stochastic process consists of a particle stochastically moving in space. In this case, the generalisation is not so trivial. Now, $\Gamma$ becomes a \textit{function} of three variables and the matrix entries of $U$ are now matrix \textit{elements}:
\begin{equation}
    \Gamma_{ij}(t), U_{ij}(t)\rightarrow\Gamma(x,y;t),\bra{x}U(t)\ket{y}\coloneq K(x,y; t)\quad ;\quad x,y\in\mathcal{C},
\end{equation}
and $U$ is now a unitary \textit{operator}, which is also a \textit{propagator}. With this, all matrix multiplications are replaced by operator concatenation and the sums over indices become integrals:
\begin{equation}
    \sum_i\Gamma_{ij}=1\rightarrow\int_\mathcal{C}\Gamma(x,y;t)d\mu_\mathcal{C}(x)=1,
\end{equation}
where $\mu_\mathcal{}$ is an appropriate measure over $\mathcal{C}$. In principle, this completes the generalisation,  but several problems arise. 
\\\\
The first one is that $\ket{x}$ is now a \textit{continuous} basis. In the position basis, we have the following normalisation condition:
\begin{equation}
    \braket{x}{y}=\delta (x-y).
\end{equation}
The next problem comes from the normalisation of $\Gamma$. To see why, let us look at the unitarity condition on $U(t)$ in this continuous case:
\begin{equation}               U(t)U^\dagger(t)=\mathbbm{1}\iff\bra{x_i}U(t)U^\dagger(t)\ket{x_j}=\braket{x_i}{x_j}=\delta(x_i-x_j)\iff\int K(x_i,x';t)K^*(x_j,x';t)dx'=\delta(x_i-x_j),
\end{equation}
where we have used the completeness relation for a continuous basis:
\begin{equation}
    \int \ket{x}\bra{x}dx=\mathbbm{1}.
\end{equation}
But this condition on the propagator $K$ implies: 
\begin{equation}
    \Gamma_{ij}=K(x_i,x_j)K^*(x_i,x_j)\implies\int\Gamma_{ij}dx_i=\int K(x_i,x';t)K^*(x_i,x';t)dx'=\delta(0),
\end{equation}
which diverges. This divergence comes from the fact that the basis used is not orthonormal in the usual sense, that is, the inner product of two basis elements is not a Kronecker delta but a \textit{Dirac} delta. Since this will happen for any \textit{continuous} basis, we see that this type of bases are not adequate for the stochastic approach.
\\\\
However, what we can do is \textit{discretise} space and work with a discrete position basis. Then, in specific examples, we could look at the limit when the discretisation step goes to zero to try to get a finite answer. Another possibility would be some kind of mathematical process, akin to regularisation in QFT, that gets rid of these divergences. However, the author is of the opinion that these problems are so fundamental --unlike in QFT, where divergences appear when trying to do \textit{perturbative} calculations-- that they seem to signal some underlying discrete nature of space (although, interestingly, not of time).
\\\\
With this, we seem to find an indication that physical space (and other possible configurations of a particle) must fundamentally be discrete in nature, which is not a new idea at all \cite{Finkelstein, Hooft, Bialynicki, Bombelli, Wolfram, Lloyd}. This hint is not totally new, since many other divergences appear throughout physics because of the non-discreteness of space\footnote{For example, we have the rather trivial example of the Newtonian or Electrostatic potential diverging for $r\to 0$. There's also the example of the divergence of self-energy of charged particles in electrodynamics or the whole family of divergences appearing in QFT.}. Despite that, time does not seem to have any limitation in this respect.

\vspace{0.5cm}
\subsection{ Fields}
Now that we have seen the problems with working with a continuous basis, we can try to do a similar analysis for a configuration space consisting of (classical) field configurations. For the sake of simplicity, we will be working with $L^2$ functions\footnote{In principle, the following analysis is valid for any $L^2$ space, \textit{i.e.} over any continuous field. However, for ease of notation, we will treat everything as one-dimensional.} . Every such function can be described using an orthonormal basis of the space by specifying the expansion coefficient for every basis element:
\begin{equation}
    \forall f(x)\in L^2\quad :\quad f=\sum_ic_i\phi_i,
\end{equation}
where $\{\phi_i\}_{i\in\mathbb{N}}$ is an orthonormal basis of $L^2$ and $c_i=(f,\phi_i)$\footnote{Here we use the notation $(\cdot,\cdot)$ for the inner product, instead of the braket notation, to distinguish the current treatment from a Hilbert space perspective} are the expansion coefficients. With this, any configuration state of the stochastic process can be specified by an infinite string of complex coefficients. Therefore, the size (cardinality) of the configuration space is:
\begin{equation}
    |\mathcal{C}|=\aleph_0\cdot2^{\aleph_0}=2^{\aleph_0}.
\end{equation}
Therefore, the configuration space is \textit{continuous}, and we have exactly the same problem as in the previous subsection. To solve this problem, however, is not sufficient to discretise space -- since that would still imply that the expansion coefficients are continuous -- but we have to also discretise the field values\footnote{To only discretise the field values wouldn't solve the problem either because, again, the expansion coefficients will be from a continuous set, and we would still have $|\mathcal{C}|=2^{\aleph_0}$.}. So we see that we have ended up in a kind of ``quantised'' field.

\vspace{0.5cm}

\subsection{ Quantum fields}
The treatment in this case is similar to classical fields. However, surprisingly, since we are already working with \textit{quantised} fields, we don't need to discretise either space nor the field values. In Quantum Field Theory (QFT), in the so-called \textit{second quantisation}\footnote{It is not a \textit{second} quantisation in the sense of going after a ``first quantisation", but it was another way of quantising a theory that historically came \textit{after} the ``first'' quantisation used in non-relativistic quantum mechanics.} or \textit{field quantisation}, we have that our states live in a \textit{Fock space}:
\begin{equation}
    \mathcal{F}=\bigoplus_{i\in\mathbb{N}}\mathcal{H}_i,
\end{equation}
where each $\mathcal{H}_i$ is the Hilbert space of $i$ particles. In turn, any particle can have a certain linear momentum $p$. Defining the vacuum, $\ket{0}$, as the state annihilated by all $a(p)$ (for all $p$), $a(p)\ket{0}=0$, we have that any state can be written as (ignoring any normalisation factors):
\begin{equation}
    \ket{\psi}=\int a^\dagger(p)\ket{0}dp+\int a^\dagger(p)a^\dagger(p)\ket{0}dp+\cdots.
\end{equation}
Since, in relativistic quantum mechanics, states of an arbitrary number of particles are allowed, we specify a state with an \textit{infinite} string of real numbers. For a state of finitely many particles, we have:
\begin{equation}
    \ket{\psi}\equiv(p_1,\cdots,p_1;p_2,\cdots,p_2;\cdots;p_n,\cdots,p_n;0,0,\cdots),
\end{equation}
where when some $p_i$ is repeated $m$ times, it means that $\ket{\psi}$ has $m$ particles with momentum $p_i$\footnote{Note that if we want to include spin in this treatment, we just have to add another entry encoding the spin in this identification.}. \\\\
Then, as before, the cardinality of the configuration space, for a quantum field, is of $\aleph_0\cdot2^\aleph_0=2^{\aleph_0}$. Therefore, we have the same problem as before of having to use a continuous basis. To solve this, however, it is enough to confine our quantum system to a \textit{box} of side $L$. In this case, the momenta particles are a discrete set:
\begin{equation}
    p_n=\frac{n\pi}{L}.
\end{equation}
Because of this, the cardinality of $\mathcal{C}$ is (drastically) lowered to $(\aleph_0)^2=\aleph_0$, so $\mathcal{C}$ is now a \textit{discrete} set, and we may use all the machinery developed in \cite{Correspondence, Theorem}. 
\\\\
With this, the state evolves according to a unitary time-evolution operator:
\begin{equation}
    \ket{\psi(t)}=e^{-i\hat{\mathcal{H}}t}\ket{\psi (0)}.
\end{equation}
Note that what evolves unitarily is the \textit{state}, and not the field operator.

\vspace{0.5cm}

\subsubsection{ Scattering process}
In a scattering process, we have an initial state at the asymptotic past, the ``in state", which we let evolve, using the time-evolution operator $U(t)$, in the presence of interactions with other particle fields, and then we compare the result with a certain final state, asymptotically into the future, the ``out state".  However, since we are dealing with \textit{interacting} fields, quantum fields no longer ``live'' in a Fock space. Nevertheless, since we treat these scattering processes perturbatively, we still work in a Fock space, in the sense that the many-particle states are used as a basis:
\begin{equation}
    \text{In state:} \quad \ket{i}\coloneq a^\dagger_i(p)\ket{0}\quad ; \quad \text{Out state:} \quad \ket{f}\coloneq a^\dagger_j(p)\ket{0}
\end{equation}
\begin{equation}
    \text{Probability:}\quad |\bra{f}U(t)\ket{i}|^2.
\end{equation}
This is exactly the expression for the individual entries of $\Gamma$. The $S$ matrix squared is then just $\Gamma$.
\\\\
We may have the concern that the in and out states are infinitely into the past and future, respectively, so that they may ``escape'' the box of length $L$ devised in the previous subsection. However, since what we actually do is calculate the scattering probability \textit{in the limit} of infinite initial and finite times, we can do the calculation for increasing sizes of $L$, and the answer should converge to some value.
\\\\
In \cite{Correspondence,Theorem}, the interference formula appearing in a measurement process is derived using the Stochastic-Quantum correspondence, without appealing to any wave-like nature of the studied entities. One may try to do a similar thing for scattering processes in QFT, where interference appears between multiple Feynman diagrams for a given perturbative order of a given process. This interference comes directly from the fact that the scattering probability is the \textit{square} of an inner product. In QFT, the unitary time-evolution operator is given by\footnote{Normally, one works in the interaction picture, so that the Hamiltonian appearing in the following equation is the interaction Hamiltonian.}:
\begin{equation}
    U(t,t_0)=\mathcal{T}\left[\exp\left(-i\int_{t_0}^t\hat{\mathcal{H}}(t')dt'\right)\right]\quad,\quad t>t_0,
\end{equation}
where $t_0$ is the initial time and $\mathcal{T}$ is the time-ordered product. Now, we can expand the exponential in its Taylor series:
\begin{equation}
    U(t,t_0)=1+\sum_{n=1}^\infty\frac{(-i)^n}{n!}\int_{t_0}^tdt_1\cdots\int_{t_0}^tdt_n\mathcal{T}(\hat{\mathcal{H}}(t_1)\cdots \hat{\mathcal{H}}(t_n)),
\end{equation}
where one does the calculation up to an order $n$ (the sum is truncated up to the n-th summand). This series expression for the time-evolution operator is also called a \textit{Dyson series}. The time-ordered products of the multiplication of many Hamiltonians contain many terms that are summed to each other. In turn, when squaring this, one gets interference terms. Therefore, we see that in this case, interference comes, ultimately, from the Born rule and from having multiple summed contributions to the total amplitude, and not from any indivisibility in the dynamics of the process-- as was the case in set-ups such as the double-slit experiment, see \cite{Correspondence}. 

\vspace{0.5cm}

\section{Practical issues}
The Stochastic-Quantum correspondence has been seen to be a major \textit{philosophical} progress in the foundation of quantum physics. However, one may ask about its practical usefulness.
\\\\
The physicist's task is normally to solve problems about how objects move (or more generally, how systems evolve with time). For this, one starts by positing the system's underlying forces (or potentials). With that, one normally gets a differential equation that is to be solved, either analytically, approximately or numerically. Crucially, one does not have the (exact) evolution of the system, but just its underlying driving forces. In the stochastic approach, however, the $\Gamma$ matrix already possesses all the information about the system's evolution (for example, the particle's trajectory), albeit possibly only stochastically and indivisibly. This means that to get the system's stochastic characterisation is to have completely solved the problem. But the issue here is that we have no way of knowing what the $\Gamma$ matrix is unless we first solve the physical problem in a traditional way! Therefore, the stochastic side of things doesn't have any \textit{practical} significance, in the sense of helping solve a physical problem.
\\\\
However, that does not mean that this Stochastic-Quantum Correspondence is completely useless. On the contrary, it yields a satisfactory unification of classical and quantum mechanics (as it is discussed in the next section) as well as providing a philosophically acceptable (and physically more palatable) foundation to quantum mechanics. Moreover, it may lead to generalisations of quantum mechanics (like quantum gravity) as well as potential applications to other fields of physics (like quantum computing). 
\\\\
As said above, $\Gamma$ has the information of the solution to the physical problem at hand. However, this information is not complete, and for the same reason that it does not \textit{uniquely} correspond to a certain Hamiltonian and basis: the modulus square relating the $\Gamma$ matrix to the unitary operator $U$. Specifically, this correspondence hides any phase, so that you can only get every component of $U$, from $\Gamma$, up to a phase. This is the ``Schur-Hadamard gauge'' described in \cite{Correspondence}. Indeed, this phase can be almost all that matters in a quantum-mechanical problem. For example, when dealing with the eigenvectors of the Hamiltonian operator, the time evolution is given by an overall phase to the state:
\begin{equation}
    \ket{\psi_n(t)}=e^{-iE_nt}\ket{\psi_n(0)}.
\end{equation}
Actually, the time evolution is uniquely given by the Hamiltonian, and what the problem actually consists of is finding a set of solutions at the initial time $t=0$. But from the $\Gamma$ matrix, evaluated at $t=0$ where $U(0)=\mathbbm{1}$, we only get the \textit{components}, in the configuration basis, of the state. The issue is, however, that the component of an element of the configuration basis is trivial:
\begin{equation}
    \braket{i}{j}=\delta_{ij}.
\end{equation}
Therefore, we don't get any information about the initial-time solution, only that the basis is orthonormal. So, all in all, we see that the stochastic approach already gives us the exact evolution of the system, without giving us any hope of solving the problem for ourselves. Moreover, it may even hide all the relevant information about the problem (the relative phases).

\vspace{0.5cm}

\section{ A unification of classical and quantum mechanics}
In quantum mechanics, the so-called ``classical limit'' has always been an ongoing discussion among scholars \cite{Schlosshauer, Joos, Bacciagaluppi, Schlosshauer2, Zurek2}.  In particular, what seems more troubling is the huge ontological differences between quantum mechanics and classical physics. Specifically, in classical mechanics objects ``exist'' and have definite trajectories and properties even when we don't measure them, whereas quantum particles only have definite properties when we measure them and, in the meantime, they exist in a ``superposition'' of different definite states (or, more appropriately, they exist in a new kind of state without definite properties). This philosophical issue is normally ignored, portrayed by the school of thought ``shut up and calculate". 
\\\\
On the other hand, quantum physicists have repeatedly tried to explain, or describe, classical physics starting from the more basic theory of quantum mechanics. However, the ontological differences between the two realms make such a unification impossible, since no matter how large, how many degrees of freedom or how correlated systems are, they won't go from having no definite properties between measurements to constantly having them, which is the world-picture we have in classical physics.
\\\\
In this regard, the stochastic point of view has much to say. Now, a quantum system as well as a classical system have the same ontological background: particles (or, more generally, systems) changing between different definite configurations (or states). The only difference is that quantum systems evolve according to stochastically indivisible dynamics, whereas classical systems do so deterministically. This change from indivisible stochastic dynamics to determinism is briefly touched upon in \cite{Correspondence} (section IV.F) and here it is explained more thoroughly. 
\\\
The important feature that bridges between the two theories is the existence of \textit{division events}. A division event is a point in time where the dynamics is momentarily no longer indivisible. That is, we can start the evolution from this division event up to later times. As we saw in Section 2 in the context of the superposition principle and in \cite{Correspondence} (section IV.D), interferences appear whenever an indivisible process is measured in the midst of it. On the other hand, if we measure it right when a division event occurs, this interference will disappear, and the measurement will have as an outcome the classical expected result.

\vspace{0.5cm}

\subsection{Measurement devices and the environment}
Following the treatment in \cite{Correspondence}, we start with a quantum system. Then the system interacts with the environment, which means that, after the interaction has taken place,  the system's configuration is \textit{perfectly} correlated to the configuration of the environment:
\begin{equation}
    p(i,e;t)=p(i;t)\delta_{e,e'(i)},
\end{equation}
where $p(i;t)$ is the probability of the system being in configuration $i$, $e$ is the configuration of the environment after the interaction, and $e'(i)$ is the environment configuration \textit{when the system is in configuration $i$} (this is where the \textit{perfect} correlation comes in). Right after the interaction, which ends at time $t_0$, the system and the environment evolve independently, \textit{i.e.} their joint wavefunction factorises into the two separate evolutions:
\begin{equation}
    \braket{i, e}{\Psi}=\braket{i}{\psi}\delta_{e, e'(i) }, 
\end{equation}
where $\ket{\Psi}$ is the wavefunction of the overall system plus environment and $\ket{\psi}$ is the wavefunction of the system alone. Then, the wavefunction for later times is given by:
\begin{equation}
    \braket{i, e}{\Psi}=\sum_{i'} \bra{i}U^\mathcal{S}(t_0\rightarrow t)\ket{i'}\braket{i'}{\psi(t_0)}\bra{e}U^\mathcal{E}(t_0\rightarrow t)\ket{e'(i')},  
\end{equation}
where the superscript $\mathcal{S}$ refers to the system alone and $\mathcal{E}$ to the environment. Then, marginalising over the configurations $e$ of the environment, and remembering that the probability is given by the modulus square of the wavefunction (Born's rule), one gets that the probability for the system alone is given by:
\begin{equation}
\begin{split}
    p(i;t)&=\sum_{i'_1,i'_2} \overline{\bra{i}U^\mathcal{S}(t_0\rightarrow t)\ket{i_1'}\braket{i_1'}{\psi(t_0)}}\bra{i}U^\mathcal{S}(t_0\rightarrow t)\ket{i_2'}\braket{i_2'}{\psi(t_0)}\\
    \sum_e&\overline{\bra{e}U^\mathcal{E}(t_0\rightarrow t)\ket{e'(i'_1)}}\bra{e}U^\mathcal{E}(t_0\rightarrow t)\ket{e'(i'_2)}
\end{split}
\end{equation}
Then, by unitarity, the sum over $e$ gives $\delta_{e(i'_1),e(i'_2)}$. Here, in \cite{Correspondence}, it is implicitly assumed that $\delta_{e'(i'_1),e'(i'_2)}=\delta_{i'_1,i'_2}$, so that you get:
\begin{equation}
     p(i;t)=\sum_{i'_1} |\bra{i}U^\mathcal{S}(t_0\rightarrow t)\ket{i_1'}|^2|\braket{i_1'}{\psi(t_0)}|^2\coloneq\sum_j\Gamma_{ij}(t_0\rightarrow t)| \braket{i}{\psi(t_0)}|^2,
\end{equation}
so that a division event appears, since the evolution can now be computed from the time $t_0$. But  $\delta_{e'(i'_1),e'(i'_2)}=\delta_{i'_1,i'_2}$ is only the case if $e'(i'_1)=e'(i'_2) \iff i'_1=i'_2$.  However, the left-to-right implication need not be the case\footnote{The right-to-left implication is always the case as long as the environment configuration is only characterised by the system configuration, which we assume here.}. In particular, it could be that the environment has the same configuration for many different configurations of the system. For example, for an environment smaller (with fewer degrees of freedom) than the system itself, there will necessarily be different configurations of the system with the same environment configuration. Arguably, an ``environment'' must have (many) more degrees of freedom than the system in question, but it may still be the case that the left-to-right implication doesn't hold for environments larger than the system.
\\\\
We are now interested in quantifying how probable it is for a very big environment (in comparison to the system) to have the implication $e'(i'_1)=e'(i'_2) \iff i'_1=i'_2$. This means that for different system configurations, different environment configurations correspond to them. So we could imagine randomly assigning an environment configuration to each system configuration and calculating the probability of repeating environment configurations --that is, having different system configurations assigned to the same environment configuration.
\\\\
Let $n$ and $m$ be the number of possible configurations of $\mathcal{S}$ and $\mathcal{E}$, respectively. Then, the total amount of possible environment configuration assignments to the $n$ system configurations is $m^n$, whereas the probability of having $n$ \textit{different} environment assignments (when drawn randomly) is $m\cdot(m-1)\cdots(m-n+1)=\frac{m!}{(m-n)!}$. Therefore, the said probability is:
\begin{equation}
    P(n, m) \equiv P(n \text{ different drawns from a pool of $m$}) =\frac{m!}{(m-n)!m^n}. 
\end{equation}
We can rewrite it as:
\begin{equation}
    P(n,m)=\frac{m}{m}\cdot\frac{m-1}{m}\cdots \frac{m-n+1}{m}=\prod_{k=0}^{n-1}\left(1-\frac{k}{m}\right).
\end{equation}
In the limit $m\gg n$:
\begin{equation}
    P(m\gg n)\approx \prod_{k=0}^{n-1}e^{-k/m}=e^{\sum_{k=0}^{n-1}-\frac{k}{m}}=e^{-\frac{n(n-1)}{2m}},
\end{equation}
where it has been used that $1-\frac{k}{m}\approx e^{-k/m}$ in this limit. Then, indeed, $P(n,m)$ tends to $1$ for large $m$. Therefore, we have found that an environment is characterised as a system with a great number of degrees of freedom compared to the studied system.
\\ If it just so happens that for every different configuration of the environment, only one system configuration is associated with it, we have that an \textit{exact} division event occurs. In this case, the ``environment'' acts as a measuring device. However, if this correspondence between environment and system configurations is not exact, the division event is only \textit{approximate}. Therefore, we see that a measuring device is any system with this correspondence, regardless of its size (as long as it has, at least, the same degrees of freedom as the studied system). On the other hand, an environment is any system that is very large compared to the studied system, such that it causes a (almost exact) division event whenever it interacts with the studied system. In particular, an environment is a type of measuring device, the difference being that, normally, as opposed to environments, measuring devices are controlled and small enough (macroscopically speaking) such that their configuration (the measurement outcome) can be easily read off.
\\\\
With this, we have the following definitions:
\begin{definition}
    An \textit{environment} is any system with many more degrees of freedom than the studied system.
\end{definition}
\begin{definition}
    A \textit{measuring device} for a certain system is another larger system whose degrees of freedom are such that every system's configuration uniquely corresponds to a measuring device's configuration.
\end{definition}
\begin{remark}
    The definition of an environment is actually not exact, and it depends on the context what exactly an environment is. Nevertheless, any macroscopic system is an environment for any quantum (subatomic) system.
\end{remark}
\begin{definition}
    A \textit{practical measuring device} is any measuring device from which we can harness information. In particular, we can easily read off its configuration.
\end{definition}
\noindent The previous definition is necessary, since from the definition of a measuring device it follows that, for instance, an exact copy of the studied system is a measuring device. However, if the studied system is microscopic, it is not useful at all as a measuring device since we can't ``read off'' the ``measuring outcome'' from it.
\\\\
From the previous discussion, we also have that:
\begin{theorem}
    Any randomly chosen environment causes an almost exact division event on the studied system.
\end{theorem}
\begin{remark}
    Here, ``randomly chosen'' just means that the environment has not been crafted specially for the system so that the probabilistic argument previously put forward is valid.
\end{remark}
\noindent Finally, it is worth noting that the previous treatment has been done for measurements of the configuration basis, but the same can be done for the eigenvectors of any self-adjoint operator, as it is explained in \cite{Correspondence} (section V.B)

\vspace{0.5cm}

\subsection{Wavefunction collapse}
We have described what a measurement device and an environment are. The next step is the result of the interaction between the studied system and the measurement device: the wavefunction collapse.
\\\\
In standard textbook quantum mechanics, the wavefunction collapse is seen as an \textit{ad hoc}, instantaneous occurrence as a result of a measurement upon the studied system. Under the stochastic approach, this phenomenon is easily explained. The said interaction creates a division event, and therefore the $\Gamma$ matrix at a time $t$ after the interaction, ending at $t_0$, can be expressed as:
\begin{equation}
    \Gamma(t>t_0)=\Gamma(t_0\rightarrow t)\Gamma(t_0)\implies \Gamma(t>t_0)=\sum_k\Gamma_{ik}(t_0\rightarrow t)|\bra{k}\psi(t_0)\ket{i}|^2,
\end{equation}
where we have used that $\ket{\psi}=U(t)\ket{j}$ and $\Gamma_{ij}(t_0)=|\bra{i}U(t)\ket{j}|^2$. Then, if we can access the information about the state of the measuring device at time $t_0$ (the measurement outcome), we know the configuration of the system at that time. Therefore, the wavefunction for later times is:
\begin{equation}
    \ket{\psi (t>t_0)}=U(t_0\rightarrow t)\ket{\psi(t_0)}.
\end{equation}
Then, the wavefunction has ``collapsed'' to the state $\ket{\psi(t_0)}$, which is an eigenstate of the measured operator. This ``collapse'' is very different from the one in a textbook treatment. That is because, in the stochastic approach, the system is \textit{always} in a definite configuration, and the collapse just comes from the fact that we can directly harness information from the measuring device (as it is mentioned in \cite{Correspondence}). In other words, the wavefunction collapse is just a consequence of conditional probability. This is as prosaic an occurrence as updating the probability distribution given the distribution at some time $t_0$, as is done in Bayesian probability theory.
\\\\
Under this view, it is true, in a sense, that the wavefunction collapse indeed depends on us ``watching'' the system (making a measurement). That is because the wavefunction collapse is a result of \textit{conditional} probability, which depends on the knowledge of the person studying. Specifically, as well as probability, the wavefunction depends on our knowledge of the system. This is, however, not a mystical occurrence, since, as opposed to textbook quantum mechanics, the ontology of the studied system doesn't change by measuring it: what changes are the probabilities assigned to it. In particular, if we have a system interacting with an environment, but we cannot recover any information from the environment (because, for example, it is too complicated for our analyses), we won't have any information about the state of the system and, therefore, no collapse will happen. On the other hand, if we use a measuring device, we \textit{will} be able to recover information from it (and, then, also from the system), causing our probability assignments to change and, with it, a wavefunction collapse. Therefore, another difference between a measuring device and an environment is the ability to harness information from it, or, in other words, its entropy. That entropy classification is also in accordance with the number of degrees of freedom discussed above.
\\\\
The key difference between textbook quantum mechanics and the stochastic approach is that in the former, the wavefunction is all there is to a system --is independent of the outer world--, while in the latter, the wavefunction represents our knowledge about the system --the probabilities \textit{we} assign to it given what we know about the system. This does not mean, however, that the wavefunction depends entirely on us, since it ultimately stems from the dynamical law encoded in $\Gamma$.  On the other hand, the wavefunction does depend on which system we are considering, and that does depend on us.
\\\\
With this, we see that the so-called ``measurement problem'' is solved in a very elegant way, by just changing the ontology (the physical picture) of the theory, from the wavefunction being a fundamental object to systems with definite configurations stochastically (and indivisibly) evolving with time.
\vspace{0.5cm}

\subsection{Macroscopic systems and the Ehrenfest equation}
Returning to the classical limit, we see that, essentially, we recover a classical, deterministic system whenever the studied system is continually interacting with the environment. That is because in that case, we have that division events continually happen, leading to divisible dynamics and continual wave-function collapse, as described in the previous subsection.
\\\\
In particular, if a \textit{microscopic} system is continually interacting with an environment, if we go and measure it, we will find that interferences disappear and coherence is lost--- in accordance with the effect known as \textit{decoherence}. If, on the other hand, the system is macroscopic, that means that it is automatically continually interacting with an environment, since even part of itself can be regarded as an environment. 
\\\\
However, what is missing is to recover the underlying classical laws for a classical system. This has been done, but in the other way around, using classical physics as the fundamental theory \cite{Nelson, Nelson2, Nelson3}. That is, the Schrödinger equation has been derived starting from Newtonian physics. However, here the goal is exactly the opposite: to regard quantum (stochastic) dynamics as fundamental and to derive classical mechanics from it. To this end, let us consider a system of $N$ particles interacting with each other. If there is a \textit{cohesive} force between them, even if they evolve stochastically, they will tend to clump and therefore form a single body. To illustrate this, we can regard a stochastic system whose configuration space is the space of all positions of the centre of mass of the system. To model a cohesive interaction between the constituent parts, we may suppose that each particle's position is an independent random variable with zero mean\footnote{By moving to the CM's reference frame, the mean is zero, whereas in general, it will be some value $\mu$ that depends on time.} and some fixed variance. Then, the expectation value of the centre of mass is:
\begin{equation}
    X_{CM}\coloneq \frac{1}{N}\sum_i X_i\quad ;\quad \langle X_{CM}\rangle=\frac{1}{N}\sum_i \langle X_i\rangle=0,
\end{equation}
where each $X_i$ is the position of the $i$-th particle and all the particles are considered to be identical. Then, indeed, the fact that each particle motion is centered around the origin--- since its expectation is zero -- roughly models an attractive (cohesive) force to the centre of mass. Then, we can calculate the variance to get:
\begin{equation}
    \sigma_{CM}\coloneq \langle X^2_{CM}\rangle-\langle X_{CM}\rangle^2=\frac{1}{N^2}\langle \sum_{i,j}X_i X_j\rangle=\frac{1}{N^2} \sum_i\langle X^2_{i}\rangle=\frac{\sigma_0}{N},
\end{equation}
where we have used that each particle is independent of the others (and hence $ \langle X_iX_j\rangle=\delta_{ij} \langle X^2_i\rangle$) and that each particle has a variance of $\sigma_0$. Therefore, we see that for an increasing number of particles, the variance decreases, meaning that the system indeed forms a single identity. This description is, however, inherently macroscopic, since we are considering the centre of mass. To have a derivation from microscopic dynamics, let's suppose that the configuration space of the stochastic system is the space of all possible individual positions of each particle:
\begin{equation}
    X\coloneq\{\vec{X}_1, \vec{X}_2,\cdots,\vec{X}_N\}.
\end{equation}
If the particles are indistinguishable, the probability of a certain configuration is:
\begin{equation}
    p(X)=N!\prod_{i=1}^Np(\vec{X_i}),
\end{equation}
where each $p(\vec{X}_i)$ is the probability that a particle is at position $\vec{X_i}$. Then, the expectation value is:
\begin{equation}
    \langle X\rangle=N!\int_{\mathcal{C}} \vec{X_i}\prod_{i=1}^Np(\vec{X_i})d\mu_\mathcal{C}, 
\end{equation}
where the integral is over all of possible positions of every particle. If we assume that the probability $p(\vec{X_i})$ only depends on the distance to the origin, the sum will vanish since every term with $\vec{X_i}$ will cancel with its opposite term with $-\vec{X_i}$. 
\\\\
In the case of vector random variables, there is not a single notion of variance (it is actually encoded in the \textit{covariance matrix}), but a good measure is the sum of the individual variances:
\begin{equation}
    \sigma_X\coloneq \sum_i \sigma_i=N\cdot \left(\langle \vec{X}_i^2\rangle-\langle \vec{X}_i\rangle^2\right)=N\langle r^2_i\rangle,
\end{equation}
where $r_i$ is the distance to the origin. For the calculation of its expectation value, we need a specific probability distribution. For simplicity's sake, let $p(\vec{X_i})=C\frac{e^{-ar_i^2}}{r_i}$:
\begin{equation}
    \langle r^2_i\rangle=C\int_0^\infty r^2\frac{e^{-ar^2}}{r}dr=\frac{C}{2a}.
\end{equation}
To model that the interaction increases with $N$ (more particles involved means more attraction), we can define $a\coloneq a_0\cdot N^m$. With this, the variance is:
\begin{equation}
   \sigma_X=\frac{C}{2N^{m-1}}. 
\end{equation}
If $m>1$, the interaction is strong enough, and the variance decreases with $N$, meaning that the system behaves more like a single entity for more and more particles.
\\\\
Now that we have seen some simplified models, we can go on to prove a more general statement regarding the classical limit whenever a large system of stochastically evolving particles is involved. First of all, we have to define some previously used notions precisely: 
\begin{definition}
\label{Cohesive}
    A system of $N$ particles is under a \textit{cohesive force} whenever the particle coordinates, relative to the centre of mass, have a covariance that tends to $0$ for the limit $N\to\infty$. Specifically, let $s_i$ be such a relative coordinate for the $i$-th particle: $s_i=r_i-R_{CM}$. Then:
    \begin{equation}
        \lim_{N\to\infty}\langle s_is_j\rangle=0.
    \end{equation}
\end{definition}
\noindent  Intuitively, this definition means that the particles' movement becomes less random whenever the number of particles increases, since a higher number of total particles means a greater attractive force to the centre of mass and therefore less random movement about it. That is because the covariance of the system tends to $0$ for large $N$, meaning that their movement is less random, and this loss of randomness due to vanishing covariance is identified as the effect of a cohesive force linking the particles together.
\\\\
Then, in a bounded system (\textit{i.e.}, one where the stochastic variables are bounded by some value: $|s_i|\le K\quad \forall i$) we can prove the following proposition:
\begin{proposition}
\label{proposition}
    In a bounded system of $N$ particles under a cohesive force, all central $m$-th moments ($m\ge2)$ for the relative coordinates tend to $0$ for $N\to\infty$.
\end{proposition}
\begin{proof}
    Using the boundedness of each random variable $s_i$:
    \begin{equation}
        \lim_{N\to\infty}|\langle s_{i_1}\cdots s_{i_m}\rangle|= \lim_{N\to\infty}|\langle (s_{i_1}s_{i_2})\cdots s_{i_m}\rangle|\le  \lim_{N\to\infty}K^{m-2}|\langle s_is_j\rangle|=0 \quad \forall m\ge2.
    \end{equation}
\end{proof}
\noindent This proposition will be crucial for the classical limit theorem that will be later proven.
\\\\
Another important definition to make is the following:
\begin{definition}
    A system of $N$ particles is under a short-range interaction whenever $\sum_{j=0}^\infty\langle s_i,s_j\rangle $ converges.
\end{definition}
\begin{remark}
    The sum in the previous definition is to be understood as a limit when the system tends to $N\to\infty$, since otherwise the sum is over a finite index $j$.
\end{remark}
\begin{lemma}
    In a system under a short-range interaction, the covariance between two particles tends to $0$ when the distance between them tends to infinity. 
\end{lemma}
\begin{proof}
    Let us order the system of particles in increasing distance to the origin \footnote{If two of them are at the same distance to the origin, just order them randomly or by any other means.}.  Then, since the sum of the short-range interaction converges, we necessarily have that:
    \begin{equation}
        \lim_{j\to\infty}|\langle s_i,s_j\rangle|=0=\lim_{d\to\infty}|\langle s_i,s_d\rangle|,
    \end{equation}
    where $d$ is a distance and $s_d$ is a particle at a distance $d$ from the particle $s_i$.
\end{proof}
\noindent This lemma gives the reason behind the name \textit{short}-range interaction, since the correlation vanishes with the distance. With this, we can prove the main theorem:
\begin{theorem}
    The centre of mass of a system  of $N$ identical particles under a cohesive, short-range, analytical interaction evolves according to Newton's second law for the expectation value:
    \begin{equation}
        M\frac{d^2\langle R_{CM}\rangle}{dt^2}=-\nabla_R V(\langle R_{CM}\rangle),
    \end{equation}
    where $V$ is the potential of the problem, the gradient is with respect to the centre of mass coordinate, and $M=Nm$ is the total mass.
\end{theorem}
\begin{proof}
    By the Ehrenfest theorem for a many-body system:
    \begin{equation}
        M\frac{d^2\langle R^{CM}_i\rangle}{dt^2}=-\langle\sum_{j=1}^N\nabla_i V( r_j)\rangle,
    \end{equation}
    where the spatial derivatives are with respect to each particle's coordinate. We can expand the potential around the centre of mass coordinate using the relative coordinates $s_i$, for fixed $i$\footnote{This is where we use the analyticity of the potential, which guarantees the convergence of the Taylor expansion to the actual function.}:
    \begin{equation}
        \nabla_iV(r_j)=\nabla_i V(s_j+R^{CM})=\nabla_iV(R^{CM})+\sum_\alpha\partial_{\alpha}\nabla_iV(R^{CM})s^\alpha_j+\sum_{\alpha,\beta}\partial_{\alpha}\partial_{\beta}\nabla_iV(R^{CM})s^\alpha_js^\beta_j+\cdots,
    \end{equation}
    where the $j$ subscript refers to the $j$-th particle and the Greek superscripts refer to the specific coordinate. Taking the expectation value on both sides and:
    \begin{equation}
         \langle\nabla_iV(r_j)\rangle=\langle\nabla_i V(s_j+R^{CM})\rangle=\langle\nabla_iV(R^{CM})\rangle+\sum_\alpha\partial_{\alpha}\nabla_iV(R^{CM})\langle s^\alpha_j\rangle+\sum_{\alpha,\beta}\partial_{\alpha}\partial_{\beta}\nabla_iV(R^{CM})\langle s^\alpha_js^\beta_j\rangle+\cdots.
    \end{equation}
    Summing over $j$, the linear term vanishes (since $\sum_i s_i=0$ for identical masses) and taking the limit for $N\to\infty$ makes the second-order terms and above vanish, because of Proposition \ref{proposition}:
    \begin{equation}
        \lim_{N\to\infty}\langle\nabla_iV(r_j)\rangle=\langle\nabla_iV(R^{CM})\rangle.
    \end{equation}
    Now we can expand again the RHS around the value $R_0\coloneq \langle R_{CM}\rangle$:
    \begin{equation}
    \begin{split}
        \nabla_iV(R^{CM})&=\nabla_iV(\langle R^{CM}\rangle)
        +\sum_\alpha \partial_\alpha\nabla_iV(\langle R^{CM}\rangle)(R_\alpha^{CM}-\langle R^{CM}_\alpha\rangle )+\\
        &\sum_{\alpha\beta} \partial_\alpha\partial_\beta\nabla_iV(\langle R^{CM}\rangle)(R_\alpha^{CM}-\langle R^{CM}_\alpha\rangle) (R_\beta^{CM}-\langle R^{CM}_\beta\rangle )+\cdots.
    \end{split}
    \end{equation}
    Taking the expectation value, the linear term identically vanishes. Because there's a cohesive force (Definition \ref{Cohesive}), in the limit for larger $N$, the second-order term also vanishes, and because of Proposition \ref{proposition}, the rest do as well. Therefore, we have:
    \begin{equation}
        \lim_{N\to\infty}\langle\nabla_iV(R^{CM})\rangle=\nabla_iV(\langle R^{CM}\rangle).
    \end{equation}
    Lastly, summing over $j$\footnote{Technically, we should have waited to do the limit for $N\to\infty$ until this point, but we would have gotten the same result.}:
    \begin{equation}
        \sum_j\nabla_iV(\langle R^{CM}\rangle)=N\nabla_iV(\langle R^{CM}\rangle)=N\frac{\partial R}{\partial r_i}\frac{\partial}{\partial R}V(\langle R^{CM}\rangle)=\frac{\partial}{\partial R}V(\langle R^{CM}\rangle).
    \end{equation}
\end{proof}
\noindent With this, we see that if a macroscopic system consisting of many particles evolves stochastically, and the attractive interaction between them is strong enough, the system will behave as a single unit. In particular, the expectation values will behave as classical variables of the entity, since, as explained in the proof of the previous theorem, the covariance and higher central moments of the centre of mass coordinate vanish in the limit of a greater number of particles.
\\\\
This is essentially the ``classical limit'' as explained in traditional, textbook quantum mechanics. However, here the difference lies in that, as described throughout this section, the stochastic approach assumes systems always have definite configurations, even for quantum-mechanical particles, making the use of the Ehrenfest equation to recover classical, deterministic dynamics truly satisfactory. 
\vspace{0.5cm}

\section{Conclusions and further work}
In this paper, we have seen how exactly the correspondence between a stochastic process and a quantum system plays out, determining how the different components of one side determine the ones on the other side. Then, the generalisation to the continuous case has been investigated, revealing some issues of normalisability with continuous bases, indicating a discrete nature of physical variables. After this, the examples of classical and quantum fields have been worked out. In the first case, it has been determined that a discretisation of space and field values is necessary, while confinement into a box is needed for the second. Finally, we have treated the measurement problem and satisfactorily recovered the classical limit. Specifically, we have found that a system with many stochastically evolving particles tends to behave like a classical body following Newton's second law. For this, the exact characterisation for an environment has been determined to be its number of degrees of freedom, whereas a low-entropy version of it is a measuring device. 
\\\\
Regarding future work, it would be very interesting to treat some other classical scenarios in QFT. This may reveal some other issues, like the one with continuous bases, which will have to be resolved and will shed more light on this field. This treatment of fields is also an introduction to the treatment of General Relativity in this Stochastic approach. Especially interesting would be the issues with relativity, where time and space characterisation changes for every frame of reference. This should have some implications for the stochastic dynamics, possibly leading to some theorems about it.


\begin{thebibliography}{99}
\bibitem{Correspondence}  Barandes, J (2023). \textit{The Stochastic-Quantum Correspondence.} \textit{Philosophy of Physics, 3(1)}. https://doi.org/10.31389/pop.186

\bibitem{Theorem} Barandes, J (2023). \textit{The Stochastic-Quantum Theorem}. arXiv preprint arXiv: 2309.03085.

\bibitem{Bell} Bell, J. S. (1964). \textit{On the Einstein Podolsky Rosen paradox}. \textbf{Physics Physique}, 1(3), 195–200.

\bibitem{Bohm} Bohm, D. (1952).\textit{ A suggested interpretation of the quantum theory in terms of “hidden” variables I \& II}. \textbf{Physical Review}, 85(2), 166–193.

\bibitem{Everett} Everett, H. (1957). “\textit{Relative state formulation of quantum mechanics}". \textbf{Reviews of Modern Physics}, 29(3), 454–462.

\bibitem{Zurek} Zurek, W. H. (2003). \textit{Decoherence, einselection, and the quantum origins of the classical}. \textbf{Reviews of Modern Physics}, 75(3), 715–775.

\bibitem{Ghirardi} Ghirardi, G. C., Rimini, A., \& Weber, T. (1986). \textit{Unified dynamics for microscopic and macroscopic systems}. \textbf{Physical Review D}, 34(2), 470.

\bibitem{Spekkens} Spekkens, R. W. (2007). \textit{Evidence for the epistemic view of quantum states: A toy theory}.\textbf{ Physical Review A}, 75(3), 032110.

\bibitem{Schlosshauer}Schlosshauer, M. (2005). \textit{Decoherence, the measurement problem, and interpretations of quantum mechanics}. 
\textbf{Reviews of Modern Physics}, 76(4), 1267–1305

\bibitem{Wallace} Wallace, D. (2012). \textit{The Emergent Multiverse: Quantum Theory According to the Everett Interpretation}. \textbf{Oxford University Press}.

\bibitem{Spekkens2} Spekkens, R. W. (2019). \textit{The Ontological Identity of Empirical Indistinguishability}. \textbf{Foundations of Physics}, 49, 1009–1030.

\bibitem{Fenyes} Fenyes, I. (1952). \textit{Eine wahrscheinlichkeitstheoretische Begründung und Interpretation der Quantenmechanik}. \textbf{Zeitschrift für Physik}, 132(1), 81–106.

\bibitem{Nelson4} Nelson, E. (1985). \textit{Quantum Fluctuations}. \textbf{Princeton University Press}.

\bibitem{Bohm2} Bohm, D., \& Hiley, B. J. (1993). \textit{The Undivided Universe: An Ontological Interpretation of Quantum Theory}. \textbf{Routledge}.

\bibitem{Ballentine2} Ballentine, L. E. (1970). \textit{The statistical interpretation of quantum mechanics}. \textbf{Reviews of Modern Physics}, 42(4), 358–381.

\bibitem{Peña} de la Peña, L., \& Cetto, A. M. (1996). \textit{The Quantum Dice: An Introduction to Stochastic Electrodynamics}. \textbf{Kluwer Academic Publishers.}


\bibitem{Dirac} Dirac, P.A.M (1930). ``The Principles of Quantum Mechanics'' (Oxf. Univ. Press).

\bibitem{Neumann} von Neumann, J. (1932). ``Mathematical Foundations of Quantum Mechanics'' (Princeton Univ. Press)

\bibitem{Hardy} Hardy, L. (2001). \textit{Quantum theory from five reasonable axioms}. arXiv:quant-ph/0101012.

\bibitem{Ariano}D’Ariano, G. M., \& Perinotti, P. (2017). Quantum Theory from First Principles: An Informational Approach. Cambridge University Press.

\bibitem{Nelson} Nelson, E. (1966). \textit{Derivation of the Schrödinger Equation from Newtonian Mechanics}. \textbf{Physical Review}, 150(4), 1079–1085.

\bibitem{Nelson2} Guerra, F., \& Morato, L. M. (1983). \textit{Quantization of dynamical systems and stochastic control theory}. \textbf{Physical Review D}, 27(8), 1774–1786.

\bibitem{Nelson3} Bacciagaluppi, G. (2020). \textit{Nelsonian Mechanics Revisited}. Studies in History and Philosophy of Modern Physics, 71, 87–100.

\bibitem{Guerra} Guerra, F., \& Ruggiero, P. (1973). \textit{New interpretation of stochastic mechanics}. \textbf{Lettere al Nuovo Cimento }(1971–1985), 8(8), 497–501.

\bibitem{Davidson} Davidson, M. (1979). \textit{A stochastic model of quantum mechanics}. \textbf{Journal of Mathematical Physics}, 20(9), 1865–1870.

\bibitem{Cufaro} Cufaro Petroni, N., \& Guerra, F. (1983). \textit{Stochastic mechanics and quantum theory}. \textbf{Foundations of Physics}, 13(3), 253–278.

\bibitem{Yasue} Yasue, K. (1981). \textit{Stochastic calculus of variations}. \textbf{Journal of Functional Analysis}, 41(3), 327–340.


\bibitem{Finkelstein} Finkelstein, D. (1969). \textit{Space-Time Code}. \textbf{Physical Review}, 184(5), 1261–1271.

\bibitem{Hooft} ’t Hooft, G. (2016). \textit{The Cellular Automaton Interpretation of Quantum Mechanics}. \textbf{Springer International Publishing}.

\bibitem{Bialynicki} Bialynicki-Birula, I. (1994). \textit{Weyl, Dirac, and Maxwell equations on a lattice as unitary cellular automata}. \textbf{Physical Review D}, 49(12), 6920–6927.

\bibitem{Bombelli} Bombelli, L., Lee, J., Meyer, D., \& Sorkin, R. (1987). \textit{Space-time as a causal set}. \textbf{Physical Review Letters}, 59(5), 521–524.

\bibitem{Wolfram} Wolfram, S. (2020). \textit{A Class of Models with the Potential to Represent Fundamental Physics}. \textbf{Complex Systems}, 29(2), 107–536.

\bibitem{Lloyd} Lloyd, S. (2006). \textit{Programming the Universe}. \textbf{Knopf.}


\bibitem{Joos} Joos, E., Zeh, H. D., Kiefer, C., Giulini, D. J. W., Kupsch, J., \& Stamatescu, I. O. (2003). ``Decoherence and the Appearance of a Classical World in Quantum Theory". \textbf{Springer}.

\bibitem{Zurek2} Zurek, W. H. (1991). \textit{Decoherence and the transition from quantum to classical}. \textbf{Physics Today}, 44(10), 36–44.

\bibitem{Schlosshauer2} Schlosshauer, M. (2007). \textit{Decoherence and the Quantum-to-Classical Transition}. \textbf{Springer}.

\bibitem{Bacciagaluppi}Bacciagaluppi, G., \& Valentini, A. (2009). ``Quantum Theory at the Crossroads: Reconsidering the 1927 Solvay Conference". Cambridge University Press.


\end{thebibliography}
\end{document}